\documentclass[a4paper,12pt]{article}

\usepackage[a4paper]{geometry}
\usepackage[dvipsnames]{xcolor}
\usepackage[utf8]{inputenc} 
\usepackage[hidelinks]{hyperref}
\usepackage{float}
\usepackage{amssymb,amsmath,amsthm}
\usepackage{amsfonts}
\usepackage{authblk}

\usepackage[draft,colorinlistoftodos,shadow, textwidth=2.5cm, textsize=scriptsize]{todonotes}
\pdfoutput=1
\usepackage{graphicx}
\usepackage{graphicx, float, tabularx, booktabs}
\usepackage{color}
\usepackage{enumerate}
\usepackage[american]{babel}
\usepackage{stackrel}

\usepackage[numbers, compress]{natbib}
\usepackage{natbib}
\newcommand{\beq}{\begin{equation}}
\newcommand{\eeq}{\end{equation}}
\newcommand{\bea}{\begin{eqnarray}}
\newcommand{\eea}{\end{eqnarray}}
\newcommand{\be}{\begin{equation}}
\newcommand{\ee}{\end{equation}}
\newcommand{\ba}{\begin{eqnarray}}
\newcommand{\ea}{\end{eqnarray}}

\usepackage{slashed}

\usepackage{amsfonts}
\usepackage{epsfig}

\usepackage{graphicx, tabularx}
\vspace{-0.1cm}
\renewcommand{\thefootnote}{\fnsymbol{footnote}}

\begin{document}
%%%%%%%%%%%%%%%%%%%%%%%%%%%%%%%%%%%%%%%%%%%%%
%

\title{Uncertainty Principles and Maximum Entropic Force}
\author[1]{Jonas Mureika\footnote{Email: jmureika@lmu.edu}}
\affil[1]{\small Department of Physics, Loyola Marymount University, Los Angeles, California, USA}
\author[2]{Elias Vagenas\footnote{Email: elias.vagenas@ku.edu.kw}}
\affil[2]{\small Department of Physics, College of Science, Kuwait University, Sabah Al Salem University City, P.O. Box 5969, Safat 13060, Shadadiya, Kuwait}

\maketitle

\begin{abstract}
\par\noindent
We consider quantum gravity corrections to the maximum entropic force that arise from several gravitational uncertainty principles. These include the Generalized Uncertainty Principle (GUP), the Extended Uncertainty Principle (EUP), the Generalized Extended Uncertainty Principle (GEUP), and the Linear-Quadratic GUP (LQGUP). We find that the modified entropic force depends on the dimensionless parameters of the uncertainty principles and, thus, on the underlying quantum gravity theory. Furthermore, the entropic force, which is quantum gravity corrected in the framework of the extended uncertainty principles, also depends on the number of Planck areas that made the ``EUP area". 
\end{abstract}

\renewcommand{\thefootnote}{\arabic{footnote}} \setcounter{footnote}{0}
%\thispagestyle{empty}
%\clearpage
%\tableofcontents
%
%
%
%%%%%%%%%%%%%%%%%%%%%%%%%%%%%%%%%%%%%%%%%%%%%
%\section{Introduction}
%%%%%%%%%%%%%%%%%%%%%%%%%%%%%%%%%%%%%%%%%%%%%
%
%
%
%
%
%
%%%%%%%%%%%%%%%%%%%%%%%%%%%%%%%%%%%%%%%%%%%%%
\section{Introduction and Maximum Force}
%%%%%%%%%%%%%%%%%%%%%%%%%%%%%%%%%%%%%%%%%%%%%
%
%
%
\par\noindent
The idea that gravity could be emergent from thermodynamics, specifically from entropy and thus information, was originally discussed in \cite{Jacobson:1995ab}, and later in \cite{Padmanabhan:2009vy,Verlinde:2010hp}. In the quantum regime, this necessitates an optimal value for related quantities, including the force. This maximum force \cite{Gibbons:2002iv, Schiller:2005eme,Barrow:2014cga,Schiller:2021jxf} is an entropic quantity whose definition is \cite{Padmanabhan:2009vy,Verlinde:2010hp}
\beq
F  = T \frac{dS}{dr_{h}} = \frac{dMc^2}{dr_{h}}~.
\label{force}
\eeq
For the Schwarzschild solution, the black hole horizon is located at $r_{h}=\frac{2GM}{c^{2}}$, and it can trivially be shown that the maximum entropic force  is of the form
\beq
F_{ent} = \frac{c^{4}}{2G}~.
\label{fmax}
\eeq
%
%
%`
\par\noindent
This establishes an upper bound on all possible forces that can be exerted in Nature. 
\par\noindent
At this point a comment is in order. In reference \cite{Gibbons:2002iv}, the expression for the entropic force contains a factor of 4 in the denominator, not 2 as written above. It should be stressed, however, that those derivations were done within the confined framework of the classical theory of general relativity of Einstein,, with no consideration for quantum effects. Indeed, Verlinde's derivation of entropic force stems from the entropy
\beq
\Delta S = \frac{2\pi k_B mc}{\hbar}\Delta x
\eeq
and so identifying the position uncertainty as (double) the horizon radius, $\Delta x = 2r_h$, the additional factor of two is recovered and the expression reduces to that in Eq.(\ref{fmax}).\\
\par\noindent
It is noteworthy that Eq.(\ref{fmax}) can also be derived from purely dimensional arguments.
Employing the Planck units $\ell_P = \sqrt{\frac{\hbar G}{c^3}}$ and  $t_P = \sqrt{\frac{\hbar G}{c^5}}$, we obtain the Planck velocity
\beq
v_P = \frac{\ell_P}{t_P}  = c
\eeq
\par\noindent
which itself is a fundamental result, in that the limiting velocity $c$ emerges from purely quantum mechanical arguments without the need for Special Relativity. The Planck acceleration is correspondingly
\beq
a_P = \frac{\ell_P}{t_P^2} = \sqrt{\frac{c^7}{\hbar G}}~.
\label{planckaccel}
\eeq
In the following letter, we use the standard Heisenberg uncertainty principle in the form
\beq
\Delta p \Delta x \geq \frac{\hbar}{2}
\eeq
\par\noindent
and, when the Planck length is employed,  the standard Heisenberg uncertainty principle is saturated as 
\bea
p_{P} \ell_{P} &=&  \frac{\hbar}{2} \nonumber\\
\left(m_{P} c\right) \ell_{P} &=& \frac{\hbar}{2}\nonumber\\
m_{P} &=&\frac{\hbar}{2\ell_P c}\nonumber
\eea%
which gives the Planck mass \footnote{In the case that the standard Heisenberg uncertainty principle was of the form $\Delta p \Delta x \geq \hbar$, then the Planck mass would have  been $m_{P}=\sqrt\frac{\hbar c}{G}$.}
\be
m_{p} = \sqrt{\frac{\hbar c}{4G}}~.
\ee
Therefore, combining this with Eq.(\ref{planckaccel}), the Planck force is of the form
\be
F_P = m_Pa_P = \frac{c^4}{2G}~.
\ee
It is evident that the Planck force, i.e., $F_{P}$, is equal to the maximum entropic force $F_{ent}$.\\
\par\noindent
At this point a number of comments are in order. 
\par\noindent
First, the physical origin of the maximum acceleration (and, thus, of the maximum force) was discussed by Caianiello \cite{Caianiello:1980iv,Caianiello:1981jq,Caianiello:1982zz} as being the result of a quantum modification of the spacetime geometry, and subsequently the object's motion along its worldline. 
In the context of special relativity, the particle's acceleration is bounded as $a(\vec{\upsilon})=\frac{(c^{2}-|\vec{\upsilon}|^{2})^{3/2}\mu^{2}}{m_{0}\hbar} \le a_{max}$ with $\hbar = \lambda \mu c$. It is  in the rest frame of the particle in which the velocity of the particle vanishes at that instant that the acceleration becomes maximum. Therefore, the maximum acceleration is considered to be related to the maximum value of the speed of a particle that cannot exceed the speed of the light \cite{Caianiello:1981jq}. 
In the case of two identical particles which interact gravitationally, if we take the mass of the particles, i.e., $\mu$, to be equal to the rest mass, i.e., $m_{0}$, then, in the center-of-mass frame, the maximum acceleration will be $a_{max}= \frac{c^{2}}{\lambda}$. This imposes a minimum value to the radius of the particles ($\sim\lambda$) which is shown to be equal to the Schwarzschild radius of the particles. Therefore, the complete collapse of the particles is avoided due to the maximum acceleration \cite{Caianiello:1981jq,Caianiello:1982zz}.
\par\noindent
Moreover, as outlined by Gibbons \cite{Gibbons:2002iv}, in any theory the temperature $T$ can be considered as energy while in a relativistic theory acceleration $a$ can be used to give an inverse length, namely $\frac{a}{c^{2}}$.  Also, in a quantum theory the temperature and time are related via the periodicity of the imaginary time, $\beta = \hbar T$. Combining the aforementioned statements yields the Unruh temperature $T=\frac{\hbar a}{2\pi c k_{B}}$.  Therefore, a theory that has a minimum length will give a maximum acceleration and this leads to a maximum temperature. 
%.
%
\par\noindent
Second, it is also worth noting that from an observation perspective, one anticipates that such maximal kinematic values could be seen in extreme gravity environments, such as black hole accretion disks. Although optical imaging of astrophysical black holes is limited, this may fall within the observable limits of the Event Horizon Telescope. Furthermore, maximum forces / accelerations may also be indirectly inferred through compact binary gravitational wave mergers in LIGO, as well as in future experiments like LISA and the Einstein Telescope.
\par\noindent
Third, one has to be very careful when applying the idea of ``maximum force" since it has been shown  in several frameworks that are beyond the classical theory of general relativity of Einstein, the upper bound of the entropic force is modified or eliminated \cite{dabrowski,Jowsey:2021ixg}. Additionally, it was shown \cite{Ong:2018xna} that when quantum gravity corrections are taken into consideration, the entropic force is weakened.
%
%
%
%
%
%%%%%%%%%%%%%%%%%%%%%%%%%%%%%%%%%%%%
\section{GUP Corrections to the Maximum Force}
\label{GUPmaxF}
%%%%%%%%%%%%%%%%%%%%%%%%%%%%%%%%%%%%
%
%
%
\par\noindent
Since the GUP introduces quantum gravitational corrections to quantum mechanics \cite{Das:2008kaa,Das:2010sj}, one can determine how this impacts \footnote{For instance, see the GUP effects on the Planck era of the universe \cite{Basilakos:2010vs}.} the maximum force by the following argument. In the standard prescription, the maximum force is calculated by entropic considerations \cite{Ong:2018xna}. We start with the GUP of the form 
\cite{Veneziano:1986zf,Gross:1987ar, Amati:1988tn, Konishi:1989wk, Maggiore:1993rv, Garay:1994en, Scardigli:1999jh}
\beq
\Delta x \Delta p \sim \frac{\hbar}{2}\left(1+ \beta\frac{ 4G}{\hbar c^{3}} \Delta p^2\right)
\eeq
\par\noindent
with $\beta$ to be the dimensionless GUP parameter. The aforesaid equation is  quadratic in $\Delta p$, whose solution is
\beq
\Delta p \sim \frac{c^{3} \Delta x}{4 \beta G }\left(1-\sqrt{1- 4\frac{\beta \hbar G}{c^{3}(\Delta x)^{2}}}\right)
\label{mom_uncertainty}
\eeq
\par\noindent
where only the negative solution has been kept in order in the limit $\beta\rightarrow 0$, namely when quantum gravity corrections are eliminated, to obtain the standard Heisenberg uncertainty principle. 
\par\noindent
Expanding Eq.(\ref{mom_uncertainty}) as a series in $\beta$ and substituting $\Delta x =r_{h}$, Eq.(\ref{mom_uncertainty}) becomes
\beq
\Delta p \sim \frac{\hbar}{2  r_{h}} + \beta \frac{\hbar^{2} G}{2 c^{3}  r_{h}^{3}} + \beta^{2} \frac{\hbar^{3} G^{2}}{c^{6}  r_{h}^{5}}
\label{GUPseries}
\eeq
\par\noindent
and, thus, we obtain
\be
\frac{dp}{dr_{h}}=-\left[ \frac{\hbar}{2  r_{h}^{2}} + \beta \frac{3 \hbar^{2} G}{2 c^{3} r_{h}^{4}} + \beta^{2} \frac{5 \hbar^{3} G^{2}}{c^{6}  r_{h}^{6}} \right]~.
\label{derivative_of_momentum}
\ee

\par\noindent
At this point a couple of comments are in order. First, here we are interested in deriving the maximum entropic force which, as we said, establishes an upper bound on all possible forces that can be exerted in Nature. The minus that appears in the RHS of the Eq.(\ref{derivative_of_momentum}) describes the fact that  when the uncertainty in momentum increases the uncertainty in position decreases (and vice versa). This has nothing to do with the magnitude of the maximum entropic force and, thus, without loss of generality, one can omit this sign in our analysis. Second, in order to obtain the maximum entropic force, one has to apply Eq.(\ref{derivative_of_momentum}) at the Planck scale, thus $p_{P}= Mc$ and the black hole horizon becomes $r_{h}=\frac{2Gm_{P}}{c^{2}}=\ell_{P}$.
\par\noindent
Therefore, employing Eq.(\ref{force}), the GUP-corrected maximum entropic force  is of the form
\be
F_{GUP}= \frac{c^4}{2G} \left(1+3\beta + 10\beta^{2}  \right)
\label{GUP_force}
\ee
which in terms of the unmodified maximum entropic force, i.e., $F_{ent}$, reads
\be
F_{GUP}= F_{ent} \left(1+3\beta + 10\beta^{2}  \right)~.
\label{GUP_force_1}
\ee
%
%
%
%
%
%
%%%%%%%%%%%%%%%%%%%%%%%%%%%%%%%%%%%
\section{EUP Corrections to the Maximum Force}
\label{EUPmaxF}
%%%%%%%%%%%%%%%%%%%%%%%%%%%%%%%%%%%
%
%
%
\par\noindent
It is known that the EUP introduces quantum gravity corrections to quantum mechanics \cite{Mureika:2018gxl}.
In this section, we investigate the impact of the EUP on the maximum entropic force. We adopt the analysis of the previous section. We start with the expression of the EUP \cite{Mureika:2018gxl}
\beq
\Delta x \Delta p \sim \frac{\hbar}{2}\left(1+\alpha \frac{\Delta x^2}{L_{*}^{2}}\right)
\eeq
\par\noindent
with $\alpha$ to be the dimensionless EUP parameter and $L_{*}$ to be some new, large (so it cannot be the Planck length), fundamental distance scale.
It is easily seen that the above equation is linear in the uncertainty of the momentum and, thus, we obtain the uncertainty in momentum to be of the form
\beq
\Delta p \sim \frac{\hbar}{2\Delta x}\left(1+\alpha \frac{\Delta x^2}{L_{*}^2}\right)~.
\label{EUP_mom_uncertainty}
\eeq
As before, we assume that the uncertainty in position is of the form $\Delta x =r_{h}$, so Eq.(\ref{EUP_mom_uncertainty}) becomes 
\beq
\Delta p \sim \frac{\hbar}{2r_{h}}+\alpha \frac{\hbar r_{h}}{2 L_{*}^{2}}~.
\eeq
and, thus, we obtain
\be
\frac{dp}{dr_{h}}=-\left[ \frac{\hbar}{2r_{h}^{2}}-\alpha \frac{\hbar }{2 L_{*}^{2}} \right]~.
\label{EUP_derivative_of_momentum}
\ee
Adopting the analysis stated in the previous section and employing Eq.(\ref{force}), the EUP-corrected maximum entropic force  is of the form
\be
F_{EUP}= \frac{c^{4}}{2G} \left(1-\alpha\frac{\ell_{P}^{2}}{L_{*}^{2}}\right)
\label{EUP_force}
\ee
which in terms of the unmodified maximum entropic force, i.e., $F_{ent}$, reads
\be
F_{EUP}= F_{ent} \left(1-\alpha\frac{\ell_{P}^{2}}{L_{*}^{2}}\right)
\label{EUP_force_1}
\ee
At this point, it is noteworthy that the  EUP-corrected maximum entropic force, i.e., Eq.(\ref{EUP_force_1}), can be expressed in terms of the number of Planck areas that make up the cosmologically-fundamental ``EUP area'' $L_*^2$,  i.e.,
\be
F_{EUP}= F_{ent} \left(1-\frac{\alpha}{N}\right)~~,~~N = \frac{L_*^2}{\ell_P^2}\label{EUP_force_2}~.
\ee
This suggests a holographically-inspired correction similar to Dvali and Gomez's quantum-N black hole portrait \cite{Dvali:2011aa}.
%
%%%%%%%%%%%%%%%%%%%%%%%%%%%%%%%%%%%%
\section{GEUP Corrections to the Maximum Force}
\label{GEUPmaxF}
%%%%%%%%%%%%%%%%%%%%%%%%%%%%%%%%%%%%

We now shift focus to the Generalized Extended Uncertainty Principle \cite{Kempf:1994su}, which is written as the generic combination of the GUP and EUP
\be
\Delta x \Delta p \sim \frac{\hbar}{2}\left(1+\frac{4\beta G}{\hbar c^3} \Delta p^2 + \alpha \frac{\Delta x^2}{L_{*}^2}\right)~.
\label{geup}
\ee
As previously noted \cite{bcjm,Casadio:2025sjp}, the above equation can be interpreted as a quadratic equation in $\Delta p$, with  solution of the form
\be
\Delta p \sim \frac{\Delta x c^3}{4\beta G}\left(1\pm\sqrt{1-\frac{4\beta G \hbar}{c^3}\left(\frac{1}{\Delta x^2}+\frac{\alpha}{L_{*}^2}\right)}\right)~.
\label{geupdp}
\ee
\par\noindent
Choosing the negative solution and expanding as a series as in Section~\ref{GUPmaxF}, and keeping terms up to second order in both $\alpha$ and $\beta$, we find
\be
\Delta p \sim \frac{\hbar}{2 r_h}+\alpha\frac{\hbar r_{h}}{2 L_{*}^2}+ \beta\frac{ G \hbar^2}{2  c^3 r_h^3}+ \alpha\beta\frac{ G \hbar^2}{ c^3 r_h  L_{*}^2}+ \alpha^2\beta\frac{G\hbar^2 r_h}{2c^3 L_{*}^4}+ \beta^2\frac{ G^2 \hbar^3}{r_h^5 c^6}+ 3\alpha\beta^{2}\frac{G^2\hbar^3}{r_h^3c^6 L_{*}^2}~.
%+\frac{r_h \beta G \hbar^2 \alpha^4}{2 c^3 L^4}
%+\frac{\beta^2 G^2 \hbar^3}{r_h^5 c^6}+\frac{3 \beta^2 G^2 \hbar^3 \alpha^2}{r_h^{3} c^6 L^2}+\frac{3 \beta^2 G^2 \hbar^3 \alpha^4}{r_h c^6 L^4}+\frac{r_h \beta^2 G^2 \hbar^3 \alpha^6}{c^6 L^6}
\label{geupdp}
\ee
This yields
\be
\frac{dp}{dr_h} = -\left[ \frac{\hbar}{2 r_h^2}- \alpha\frac{\hbar}{2 L_{*}^2}+3\beta\frac{ G \hbar^2}{2 r_h^4 c^3}+ \alpha\beta \frac{ G \hbar^2}{r_h^2 c^3 L_{*}^2}- \alpha^2\beta\frac{ G\hbar^2}{2c^3 L_{*}^4} + 5\beta^2\frac{ G^2 \hbar^3}{r_h^6 c^6} + 9\alpha\beta^2\frac{G^2\hbar^3}{r_h^4c^6 L_{*}^2} \right]
%\frac{dp}{dr_h} \sim -\frac{\hbar}{2r_h^2}+\frac{\hbar \alpha^2}{2L^2}-\frac{3\beta G\hbar^2}{2r_h^4c^3}-\frac{\alpha^2\beta G\hbar^2}{r_h^2 c^3 L^2}
\label{geupdpdrh}
\ee
from which the maximum force can be extracted as
\be
F_{\rm GEUP} =  
\frac{c^4}{2G} \left(1+3\beta + 10\beta^{2} 
- \alpha \frac{\ell_{P}^{2}}{L_{*}^2}
+ 2\alpha\beta\frac{\ell_{P}^{2}}{L_{*}^2} +18\alpha\beta^2\frac{\ell_{P}^{2}}{L_{*}^2}
-\alpha^2\beta\frac{\ell_{P}^{4}}{L_{*}^4}\right)~.
\label{GEUP_force}
\ee
which in terms of the unmodified maximum entropic force, i.e., $F_{ent}$, reads
\be
F_{\rm GEUP} =  
F_{ent} \left(1+3\beta + 10\beta^{2} 
- \alpha \frac{\ell_{P}^{2}}{L_{*}^2}
+ 2\alpha\beta\frac{\ell_{P}^{2}}{L_{*}^2} +18\alpha\beta^2\frac{\ell_{P}^{2}}{L_{*}^2}
-\alpha^2\beta\frac{\ell_{P}^{4}}{L_{*}^4}\right)~.
\label{GEUP_force_1}
\ee
\par\noindent
At this point a number of comments are in order. First, the GEUP maximum entropic force, i.e., Eq.(\ref{GEUP_force_1}), contains that exact GUP corrections from Eq.(\ref{GUP_force}), as well as the EUP correction from Eq.(\ref{EUP_force}). Second, there are cross terms combining both GUP and EUP parameters. Third, the EUP corrections again include the dimensionless ratio $\ell_P^2/L_*^2$, which suggests a possible connection to holography or the Bekenstein bound.
%That is, $N=L_*^2/\ell_P^2$ is the number of Planck areas in the presumably cosmologically-fundamental area $L_*^2$.

%
%
%
%
%%%%%%%%%%%%%%%%%%%%%%%%%%%%%%%%%%%%
\section{Linear-Quadratic GUP Corrections to the Maximum Force}
\label{QGUPmaxF}
%%%%%%%%%%%%%%%%%%%%%%%%%%%%%%%%%%%%
%
%
%
\par\noindent
In this section we utilize a GUP with linear and quadratic term  in momentum (LQGUP) of the form \cite{Ali:2009zq,Ali:2010yn,Ali:2011fa}
\be
\Delta x \Delta p \sim \frac{\hbar}{2}\left(1 - \beta \frac{\Delta p}{m_{p} c} +4\beta^{2} \frac{\Delta p^{2}}{m_{p}^{2}  c^{2}}\right)
\label{lqgup}
\ee
\par\noindent 
with the LQGUP parameter, i.e., $\alpha$, to be dimensionless and different from the dimensionless EUP parameter that appears in Sections 
\ref{EUPmaxF} and \ref{GEUPmaxF}.
It is obvious that Eq.(\ref{lqgup}) as a quadratic equation in $\Delta p$ can be solved and its solution reads
\be
\Delta p =
\frac{\left( \beta\frac{\hbar}{2}+ \Delta x\right)\pm \sqrt{\left( \beta\frac{\hbar}{2}+ \Delta x\right)^{2}-4\frac{\hbar}{2}(2\hbar\beta^{2})}}{4\hbar \beta^{2}}~.
\label{lqgupdp}
\ee
Adopting the analysis of the previous sections, we obtain
\be
\Delta p = \frac{\hbar}{2r_{h}}+ \beta^{2}\frac{\hbar^{3}}{2 m_{p}^{2}c^{2}r_{h}^3}
\label{lqgupdp_taylor}
\ee
which yields
\be
\frac{dp}{dr_{h}}=-\left[ \frac{\hbar}{2r_{h}^{2}}+3\beta^{2}\frac{\hbar^{3}}{2 m_{p}^{2}c^{2}r_{h}^{4}}   \right]~.
\label{lqgupdpdr}
\ee
Therefore, the LQGUP maximum entropic force is of the form
\be
F_{LQGUP}=\frac{c^{4}}{2G} \left( 1+12\beta^{2} \right)
\label{lqgup_force}
\ee
which in terms of the unmodified maximum entropic force, i.e., $F_{ent}$, becomes
\be
F_{LQGUP}=F_{ent} \left( 1+12\beta^{2} \right)~.
\label{lqgup_force_1}
\ee
%
%%%%%%%%%%%%%%%%%%%%%%%%%%%%%%%%%
\section*{Conclusions}
In this paper, we have considered ultraviolet corrections to the maximum entropic force (and by proxy, the maximum acceleration \cite{GallegoTorrome:2018qfe}) introduced by various higher-order forms of the uncertainty principle. We show, as expected, the corrections depend on  the associated dimensionless parameter $\beta$ for the GUP and related forms, and also $\alpha$ for the EUP. Perhaps unsurprisingly, the correction due to the EUP mimics the form of the EUP correction to Einstein gravity introduced in \cite{Mureika:2018gxl}, albeit with an overall sign difference in the correction. It is also interesting to note that the correction term $\ell_P^2/L_*^2$ is related to associated Hawking temperature for black holes at the crossover mass scale $R_S \sim L_*$, which gives $M_{\rm crit} \sim L_*/\ell_P^2$.

Although the values of $\alpha$ and $\beta$ are not known, there have been arguments made to suggest that at least $\beta$ should be of order unity \cite{Amati:1988tn,Konishi:1989wk,Gross:1987kza,Capozziello:1999wx}. In this case, the upper bound for corrections to the maximum force are very large for the GUP-related quantities. However, if the value of $\beta \sim 0.1$, the corrections themselves will be of order unity.

%\tcr{JM Question: Is this better in the conclusions as potential experiemtal constraints, or alternatively in the introduction as motivation? Also, can we suggest something like inflation / dark energy as a constraint on the parameters? Maybe this could provide a new link between quantum gravity and $\Lambda$. }

%%%%%%%%%%%%%%%%%%%%%%%%%%%%%%%%
%
%
%
%
%%%%%%%%%%%%%%%%%%%%%%%%%%%%%%%%
\section*{Acknowledgements}
JM thanks the Department of Physics and College of Science at Kuwait University for their generous and welcoming hospitality, during the initial stages of this research.

%%%%%%%%%%%%%%%%%%%%%%%%%%%%%%%%
%
%
%
%\bibliographystyle{inspires}
%\bibliographystyle{physlett}	% (uses file "inspires" "inspires_t_n" "plain.bst", "apsrev.bst", "apsrmp.bst", "unsrt.bst" "JHEP.bst" "unsrtnat.bst" iopart-num, habbrv, hacm, halpha, hapalike, hieeetr )
%\bibliographystyle{apsrev4-1}
%\bibliography{jmrefs}
%expects file "myrefs.bib"
%\cite{Mureika:2018gxl}
%
%
%
%
%
%
%
%%%%%%%%%%%%%%%%%%%%%%%%%%%

%%%%%%%%%%%%%%%%%%%%%%%%%%%
%
%
%
%
%
%
%
\end{document}